\documentclass[cits]{PoS}

\newcommand{\mean}[1]{\langle #1 \rangle} 

\title{Multiparticle correlations and momentum conservation in nucleus-nucleus 
  collisions}

\ShortTitle{Multiparticle correlations and momentum conservation}

\author{\speaker{Nicolas Borghini}\\
        Fakult\"at f\"ur Physik, Universit\"at Bielefeld, 
        Postfach 100131, D-33501 Bielefeld, Germany\\
        E-mail: \email{borghini@physik.uni-bielefeld.de}}

\abstract{Particle correlations are very actively studied in heavy-ion 
  collisions at ultra-relativistic energies. 
  Here, an attempt is made at determining a proper reference for such studies, 
  by taking properly into account the multiparticle correlations induced by the 
  conservation of total momentum in the collisions.}  

\FullConference{High-pT physics at LHC\\
		March 23-27 2007\\
		University of Jyv\"askyl\"a, Jyv\"askyl\"a, Finland}

\begin{document}

\section{Introduction}
\label{s:intro}

High-energy collisions of heavy nuclei result in the emission of up to several 
thousands of particles. 
Such large numbers reflect the complexity of the collisions; yet, they allow for
a wealth of different tools to characterize the processes that take place, in 
particular to assess the properties of the medium that is created.
Among the possible approaches, the presence of these many particles calls in a 
natural way for statistical descriptions, as well as for observables involving 
several particles. 
Various studies are thus devoted to extracting correlations between pairs, 
triplets, and more generally $n$-tuples of particles. 

The purposes of these investigations are manifold, as are the underlying physics
pictures: 
To mention only correlations in momentum space, there are studies focusing on 
identical particles that are close in momentum (and in position) space, 
searching for evidence of (anti)symmetrization of the corresponding 
wave-function~\cite{Lisa:2005dd}. 
Other, closely related studies deal with pairs of non-identical particles with 
close velocities, to estimate the difference between their emission 
times~\cite{Lednicky:1995vk}. 
Another phenomenon investigated through studies of the correlations between 
emitted particles is the anisotropy in the transverse plane of the emission 
pattern induced by the finite impact parameter between the nuclei (``anisotropic
flow'')~\cite{Tang:2007kv}. 
The extended transverse-momentum range that became available with collisions 
at RHIC allowed novel investigations, involving particles with high transverse 
momenta $p_T$, with a view to studying jets. 
A first kind of such study is that of correlations in azimuth between pairs of 
particles with both high transverse momenta~\cite{Adler:2002tq}, to look for 
the presence of structures, close or away in azimuth to a high-$p_T$ reference 
(``trigger'') particle, that resemble the jets seen in $pp$ collisions.
The next step consists of studies of correlations between a high-$p_T$ trigger 
and low-$p_T$ ``associated'' particles~\cite{Adams:2005ph}, so as to 
characterize the response of the created medium to the propagation of a 
high-$p_T$ parton (or, less plausibly, hadron).

Whatever the specific aim of a correlation study, it boils down to a simple 
principle, namely to try to identify the trace of some genuine dynamical effect 
in the joint two-, three-, $M$-particle distributions.
To accomplish this, one needs to determine properly what the expectation would 
be for these distributions in the absence of such a dynamical effect. 
Now, such a ``no-dynamics'' $M$-particle distribution is not merely the product 
of $M$ single-particle distributions, for there exists a trivial correlation 
between arbitrary final-state particles, due to the conservation of total 
momentum, which imposes some constraints on the joint distributions~\cite{%
  Foster:1973mt}.
The computation of these constraints will be discussed in section~\ref{%
  s:mom-cons&cums}, using the general formalism from probability theory 
introduced in section~\ref{s:proba}.
In section~\ref{s:discussion}, I shall further discuss the meaning of this 
ever-present correlation due to global momentum conservation in the collision, 
and speculate on possible ways to take it into account in correlation studies.

\section{Probability distributions and cumulants}
\label{s:proba}

Consider a collision with a total of $N$ particles in the final state\footnote{%
  $N$ includes {\em all\/} particles, including the non-detected ones.} 
(throughout this paper, $N$ is assumed to be large).
The basic observable in studies of the momentum correlations between $M$ 
particles among the $N$ is the joint distribution 
${\rm d}^M\!{\cal N} /\, {\rm d}{\bf p}_1\!\cdots {\rm d}{\bf p}_M$.
To avoid normalization issues, it is more convenient to consider the joint
$M$-particle {\em probability\/} distribution $f({\bf p}_1,\ldots,{\bf p}_M)$, 
which is by definition normalized to unity, and therefore (roughly) independent 
of the system size. 

By definition, $M$ particles with momenta ${\bf p}_1,\ldots,{\bf p}_M$ are 
statistically independent from each other if and only if the corresponding 
joint probability distribution can be factorized into the product of the $M$ 
single-particle probability distributions: 
$f({\bf p}_1,\ldots,{\bf p}_M)=f({\bf p}_1)\cdots f({\bf p}_M)$.
Reciprocally, if they are not independent, this factorization no longer holds 
and the joint probability distribution involves further terms: the joint 
probability distribution can be expanded into a sum over all products of 
{\em cumulants\/} corresponding to distinct partitions of the $M$ particles. 
For instance, at the two- and three-particle levels:
\begin{eqnarray}
f({\bf p}_1,{\bf p}_2) & = & 
  f_c({\bf p}_1) f_c({\bf p}_2) + f_c({\bf p}_1,{\bf p}_2), \label{2-cum}\\
f({\bf p}_1,{\bf p}_2,{\bf p}_3) & = & 
  f_c({\bf p}_1) f_c({\bf p}_2) f_c({\bf p}_3) + 
  f_c({\bf p}_1) f_c({\bf p}_2,{\bf p}_3) + 
  f_c({\bf p}_2) f_c({\bf p}_1,{\bf p}_3) + 
  f_c({\bf p}_3) f_c({\bf p}_1,{\bf p}_2) \nonumber\\
 & & \hspace{3.02cm} +\, f_c({\bf p}_1,{\bf p}_2,{\bf p}_3), \label{3-cum}
\end{eqnarray}
where the single-particle cumulant is equal to the single-particle probability 
function, $f_c({\bf p})=f({\bf p})$.
The $M$-particle cumulant $f_c({\bf p}_1,\ldots,{\bf p}_M)$ corresponds to the 
``genuine'' correlation between the $M$ particles. 
The physical interpretation of the cumulant expansion is straightforward: the 
joint $M$-particle distribution depends not only on the genuine $M$-particle 
correlation, but also on all the possible correlations involving subsets among 
the $M$ particles. 
As an example, think of two particles emitted exactly back-to-back with 
${\bf p}_2=-{\bf p}_1$ (as the two pions from a decaying $\rho$ meson in the 
rest frame of the latter). 
The two-particle probability distribution reads 
$f({\bf p}_1,{\bf p}_2)=f({\bf p}_1)\delta({\bf p}_1+{\bf p}_2)$ (where, in the 
$\rho\to\pi\pi$ case, the single-particle distribution $f({\bf p}_1)$ actually 
reduces to the angular distribution since $|{\bf p}_1|$ is also fixed), while 
the corresponding cumulant is obviously non-vanishing, see 
equation~(\ref{2-cum}).
Other illustrations of the difference between the distributions $f$ and the 
cumulants $f_c$ can be found in reference~\cite{Pruneau:2006gj}.

While equations~(\ref{2-cum})-(\ref{3-cum}), and so on, can be inverted one 
after the other to yield the cumulants as functions of the joint probability 
distributions, there is a more systematic way to perform the same operation. 
First, one defines a generating function of the joint multiparticle probability 
distributions:
\begin{equation}
\label{genfunc}
G(x_1,\ldots,x_N) \equiv 1 + x_1 f({\bf p}_1) + x_2 f({\bf p}_2) + \cdots + 
x_1 x_2 f({\bf p}_1,{\bf p}_2) + \cdots,
\end{equation}
and similarly for every order $M$, where $x_1, \ldots, x_N$ are auxiliary 
variables.
Given this generating function of the joint probability distributions, the 
function that generates the cumulants is simply its logarithm~\cite{vanKampen}:
\begin{equation}
\label{genfunccum}
\ln G(x_1,\ldots,x_N) \equiv x_1 f_c({\bf p}_1) + x_2 f_c({\bf p}_2) + \cdots +
  x_1 x_2 f_c({\bf p}_1,{\bf p}_2) + \cdots.
\end{equation}
Thus, the knowledge of one of these functions automatically translates into that
of the other. 
In the following section, I shall use this property to sketch the computation 
of the multiparticle cumulants due to the momentum-conservation constraint, 
starting from the expression of the joint probability distribution for $M$ 
particles. 
Moreover, I shall also use ``scaled'' cumulants 
$\bar f_c({\bf p}_1,\ldots,{\bf p}_M)\equiv f_c({\bf p}_1,\ldots,{\bf p}_M) / 
[f({\bf p}_1)\cdots f({\bf p}_M)]$.

\section{Multiparticle cumulants from momentum conservation}
\label{s:mom-cons&cums}

As stated in the introduction, the purpose is to determine how global momentum 
conservation affects the joint multiparticle probability distributions of 
final-state particles in a large-multiplicity event like a heavy-ion collision.
That is, given $N$ particles with momenta ${\bf p}_1$, \ldots, ${\bf p}_N$ 
obeying the constraint ${\bf p}_1+\cdots+{\bf p}_N={\bf 0}$, what is the 
resulting $M$-particle cumulant? 
This question was to my knowledge first addressed (albeit semi-quantitatively)
in the two-particle case in reference~\cite{Foster:1973mt}. 
A quantitative estimate of the two-particle correlation was then derived in the 
frame of some anisotropic-flow measurement~\cite{Danielewicz:1987in}, then 
independently rediscovered in the same context~\cite{Borghini:2000cm}. 
In both these cases, the computation relied on the use of the central-limit 
theorem for the distribution of the sum of $n\gg 1$ (yet $n<N$) 
{\em uncorrelated\/} momenta. 
Eventually, a general approach to compute the cumulants to arbitrary order was 
introduced in reference~\cite{Borghini:2003ur}, making use of the generating 
functions of joint probability distributions and of cumulants and of a 
saddle-point integration. 

The starting point of these calculations is the expression of the joint
$M$-particle probability distribution under the momentum-conservation 
constraint:
\begin{equation}
\label{def_Mdis}
f({\bf p}_1, \ldots, {\bf p}_M) \equiv 
\frac{\displaystyle \left(\prod_{j=1}^M F({\bf p}_j) \right) 
\int \!\delta({\bf p}_1+\cdots+{\bf p}_N) 
\prod_{j=M+1}^N \left[ F({\bf p}_j)\,{\rm d}{\bf p}_j \right]}{
\displaystyle \int \!\delta({\bf p}_1+\cdots+{\bf p}_N) 
\prod_{j=1}^N \left[ F({\bf p}_j)\,{\rm d}{\bf p}_j \right]},
\end{equation}
where $F({\bf p})$ is the single-particle probability distribution 
``unrenormalized'' for the momentum conservation constraint: 
to leading order in $1/N$, it equals the measured single-particle probability 
distribution $f({\bf p})$, yet they actually differ at the next-to-leading 
order~\cite{Borghini:2000cm}. 
In the previous calculations~\cite{Danielewicz:1987in,Borghini:2000cm,%
  Borghini:2003ur}, $F$ was assumed to be isotropic; here we shall relax this 
assumption and consider the more realistic case of collisions with anisotropic 
expansion (flow) in the transverse plane.\footnote{The recipe for extending the 
  calculation to the non-isotropic case was briefly given in reference~\cite{%
    Borghini:2003ur}, and the corresponding expression for two-particle 
  correlations can be found in reference~\cite{Chajecki:2006hn}.} 
Quite obviously, the Dirac $\delta$ in equation~(\ref{def_Mdis}) represents the 
constraint from global momentum conservation: 
in its absence, $F({\bf p})=f({\bf p})$ and the joint probability distribution 
would simply factorize into the product $F({\bf p}_1)\cdots F({\bf p}_M)$.

To derive the cumulants arising from global momentum conservation, the most 
convenient and systematic way is to introduce equation~(\ref{def_Mdis}) in the 
expression of the generating function~(\ref{genfunc}) so as to compute the 
latter~\cite{Borghini:2003ur}. 
After introducing a Fourier representation of the Dirac distribution, one finds 
that $G(x_1,\ldots,x_N)$ can be expressed as the integral over the Fourier 
conjugate variable ${\bf k}$ of the exponential of a function 
$N{\cal F}({\bf k})$, which also depends on the auxiliary variables 
$x_j$.\footnote{More precisely, ${\cal F}$ only depends on $x_j$ through 
  combinations $x_j F(p_j)/N$: this allows one to replace $x_j$ in $G$ by 
  $\bar x_j\equiv x_j F(p_j)$ --- which amounts, to leading order in $1/N$, to 
  replacing the cumulants $f_c$ by the scaled cumulants $\bar f_c$ --- and to 
  derive the scaling with $N$ of the cumulants~\cite{Borghini:2003ur}: the 
  $M$-particle cumulant scales as $1/N^{M-1}$.}
Since $N$ is supposed to be large, this integral can be performed by a 
saddle-point approximation, provided one finds the position ${\bf k}_0$ of the 
maximum. 
The key to obtaining the successive cumulants is then first to solve to a given 
order in $\bar x/N$ the equation giving ${\bf k}_0$ (a solution to the order 
$M-1$ is required for the $M$-particle cumulant):
\begin{equation}
\label{F'(k0)}
\left( \sum_{j=1}^N \frac{\bar x_j}{N} 
  \frac{{\rm e}^{{\rm i}{\bf k}_0\cdot{\bf p}_j}}
    {\mean{{\rm e}^{{\rm i}{\bf k}_0\cdot{\bf p}}}} - 1 \right) 
\mean{{\bf p}{\rm e}^{{\rm i}{\bf k}_0\cdot{\bf p}}} = 
\sum_{j=1}^N \frac{\bar x_j}{N} {\bf p}_j\,
  {\rm e}^{{\rm i}{\bf k}_0\cdot{\bf p}_j},
\end{equation}
where the angular brackets denotes a $F({\bf p})$-weighted average.
Then, one uses this expression of ${\bf k}_0$ to compute the value of 
$N{\cal F}({\bf k}_0)$:
\begin{equation}
\label{F(k)}
N{\cal F}({\bf k}_0) \equiv 
N\left(\ln \mean{{\rm e}^{{\rm i}{\bf k}_0\cdot{\bf p}}} + 
  \sum_{j=1}^N \frac{\bar x_j}{N} \frac{{\rm e}^{{\rm i}{\bf k}_0\cdot{\bf p}_j}}
    {\mean{{\rm e}^{{\rm i}{\bf k}_0\cdot{\bf p}}}} \right). 
\end{equation}
The coefficient of $\bar x_{j_1}\ldots\bar x_{j_M}$ in $N{\cal F}({\bf k}_0)$ is 
then the scaled cumulant $\bar f_c({\bf p}_1,\ldots,{\bf p}_M)$. 
For instance, one finds (we assume for the sake of brevity that 
$\mean{\bf p}={\bf 0}$, otherwise one only need replace ${\bf p}$ by 
${\bf p'} \equiv {\bf p} -\mean{\bf p}$ in the following formulas)
\begin{eqnarray}
\label{c2}
\bar f_c({\bf p}_1, {\bf p}_2) &=& \displaystyle 
-\frac{p_{1,x} p_{2,x}}{N\mean{p_x^2}} - \frac{p_{1,y} p_{2,y}}{N\mean{p_y^2}} -
\frac{p_{1,z} p_{2,z}}{N\mean{p_z^2}}, \\
\bar f_c({\bf p}_1, {\bf p}_2, {\bf p}_3) & = & \displaystyle
-\!\!\!\sum_{1\leq j<k\leq 3} \left( \frac{p_{j,x} p_{k,x}}{N^2\mean{p_x^2}} + 
  \frac{p_{j,y} p_{k,y}}{N^2\mean{p_y^2}} + 
  \frac{p_{j,z} p_{k,z}}{N^2\mean{p_z^2}} \right) \cr
 & & + \displaystyle 
  \left( \frac{p_{1,x} p_{2,x}}{N^2\mean{p_x^2}} + 
    \frac{p_{1,y} p_{2,y}}{N^2\mean{p_y^2}} + 
    \frac{p_{1,z} p_{2,z}}{N^2\mean{p_z^2}} \right)\!\!
  \left( \frac{p_{1,x} p_{3,x}}{N^2\mean{p_x^2}} + 
    \frac{p_{1,y} p_{3,y}}{N^2\mean{p_y^2}} + 
    \frac{p_{1,z} p_{3,z}}{N^2\mean{p_z^2}} \right) \cr
 & & + \displaystyle 
  \left( \frac{p_{1,x} p_{2,x}}{N^2\mean{p_x^2}} + 
    \frac{p_{1,y} p_{2,y}}{N^2\mean{p_y^2}} + 
    \frac{p_{1,z} p_{2,z}}{N^2\mean{p_z^2}} \right)\!\!
  \left( \frac{p_{2,x} p_{3,x}}{N^2\mean{p_x^2}} + 
    \frac{p_{2,y} p_{3,y}}{N^2\mean{p_y^2}} + 
    \frac{p_{2,z} p_{3,z}}{N^2\mean{p_z^2}} \right) \cr
 & & + \displaystyle 
  \left( \frac{p_{1,x} p_{3,x}}{N^2\mean{p_x^2}} + 
    \frac{p_{1,y} p_{3,y}}{N^2\mean{p_y^2}} + 
    \frac{p_{1,z} p_{3,z}}{N^2\mean{p_z^2}} \right)\!\!
  \left( \frac{p_{2,x} p_{3,x}}{N^2\mean{p_x^2}} + 
    \frac{p_{2,y} p_{3,y}}{N^2\mean{p_y^2}} + 
    \frac{p_{2,z} p_{3,z}}{N^2\mean{p_z^2}} \right). \label{c3}
\end{eqnarray}
In these expressions, the $x$-, $y$- and $z$-directions are the principal axes 
that diagonalize the tensor $\mean{{\bf p}\otimes {\bf p}}$:
in practice, one axis will lie along the beam direction, one along the impact 
parameter of the nucleus-nucleus collision, which is the only preferred 
direction in the transverse plane, and the third one will be perpendicular to 
the other two. 
If $F({\bf p})$ is isotropic, $\mean{p_x^2}=\mean{p_y^2}=\mean{p_z^2}$, so that 
one recovers the formulas given in reference~\cite{Borghini:2003ur}. 

In nucleus-nucleus collisions at ultra-relativistic energies, the mean square 
momentum along the beam direction ($z$) is typically significantly larger than 
the mean square transverse components. 
As a consequence, the terms involving the $z$-component in the expressions of 
the cumulants are much smaller than those involving the other two components, 
especially when one considers final-state particles emitted close to 
mid-rapidity. 
Therefore, I shall drop these longitudinal-momentum terms from now on, which 
amounts to considering only the constraint imposed by the conservation of total
transverse momentum ${\bf p}_T$.

\section{Defining a minimally-biased background for correlation studies}
\label{s:discussion}

The behaviour of the three-particle cumulant resulting from the 
momentum-conservation constraint, equation~(\ref{c3}), was discussed in some 
detail in reference~\cite{Borghini:2006yk} in the case of an isotropic 
transverse emission of particles. 
Here, I shall now discuss further the expression of the two-particle cumulant 
and its content. 
The physical meaning of equation~(\ref{c2}) is intuitive: 
momentum conservation induces an ``anti-correlation'' between the momenta of 
final-state particles. 
This is of course reminiscent of the situation in which only two particles are 
emitted, with back-to-back equal momenta (note, however, that the calculation 
reported here cannot cover that case, since it assumes $N\gg 1$). 
Additionally, the anti-correlation is largest between particles with larger 
momenta; it is also largest in events with a smaller total multiplicity $N$.

Let me emphasize the implications of the expression of the two-particle 
cumulant. 
In collisions in which $N$ particles are in the final state, equation~(\ref{c2})
implies that the joint two-particle probability distribution reads (assuming 
first that particles are emitted isotropically, which is a fair approximation 
in the transverse plane for central collisions, so that 
$\mean{p_x^2}=\mean{p_y^2}=\mean{p_T^2}/2$):
\begin{equation}
\label{c2t}
f({{\bf p}_T}_1,{{\bf p}_T}_2) = f({{\bf p}_T}_1) f({{\bf p}_T}_2)
\left( 1 - \frac{2\,{{\bf p}_T}_1\cdot {{\bf p}_T}_2}{N\mean{p_T^2}} \right).
\end{equation}
In other words, although the particle emission is isotropic, i.e.\ 
$f({{\bf p}_T})$ depends on $p_T$ only, yet given a ``trigger'' particle with 
transverse momentum ${{\bf p}_T}_1$, the {\em conditional\/} probability to find
an ``associated''  particle with transverse momentum ${{\bf p}_T}_2$ is not 
isotropic:
\begin{equation}
\label{f(2|1)}
f({{\bf p}_T}_2 \big| {{\bf p}_T}_1) = 
\frac{f({{\bf p}_T}_1,{{\bf p}_T}_2)}{f({{\bf p}_T}_1)} = 
f({{\bf p}_T}_2)
\left( 1 - \frac{2\,{{\bf p}_T}_1\cdot {{\bf p}_T}_2}{N\mean{p_T^2}} \right) \neq
f({{\bf p}_T}_2).
\end{equation}
Thus, the conservation of transverse momentum induces a sinusoidal modulation of
the probability distribution of ``associated'' particles, with a minimum of the 
distribution along the trigger-particle momentum: 
there is a larger probability that the momentum of the associated particle 
points in the hemisphere opposite to that of the trigger particle. 
The amplitude of the modulation increases with the values of both the trigger 
momentum ${p_T}_1$ and the associated momentum ${p_T}_2$, and it decreases with 
increasing $N$.

One recognizes in equation~(\ref{f(2|1)}) the difference between 
{\em marginal\/} [$f({{\bf p}_T}_2)$] and {\em conditional\/} 
[$f({{\bf p}_T}_2 \big| {{\bf p}_T}_1)$] probability 
distributions~\cite{vanKampen}.  
Since the former is --- up to a normalization factor --- the measured 
single-particle distribution, it is most tempting to use it in studies of 
two-particle correlations;
yet one must rather use the latter once a first particle momentum ${p_T}_1$ in 
the event is fixed as a reference. 
This is not to be unexpected: by removing from the event this trigger particle, 
one obtains a collection of final-state particles (including the non-measured 
ones) which is {\em not\/} a ``valid'' event, since the sum of the transverse 
momenta of the particles does not vanish. 
Hence there is no reason why the single-particle probability distribution for 
this collection should be the same as for real events satisfying momentum 
  conservation.\footnote{In fact, from the mathematical point of view, these 
  collections of $N-1$ particles with non-vanishing total transverse momentum 
  $\sum {\bf p}_T=-{{\bf p}_T}_1$ are equivalent to events with a mean 
  first-harmonic anisotropic-flow component (directed flow) 
  $\bar v_1= -{p_T}_1/[(N-1)\mean{p_T}]$, the direction of the equivalent 
  reaction plane being that of ${{\bf p}_T}_1$. 
  Therefore, the corresponding single-particle distribution includes a 
  $2v_1\cos(\varphi-\varphi_1)$ azimuthal modulation, which is absent from the 
  original single-particle distribution of the real events, which we assumed 
  isotropic.}
Thus, in studies of two-particle correlations in which the transverse momentum
${{\bf p}_T}_1$ of one of the particles has been fixed --- thereby implicitly 
selecting a subset of the whole available sample of events --- one should 
consider the conditional probability distribution 
$f({{\bf p}_T}_2 \big| {{\bf p}_T}_1)$ of equation~(\ref{f(2|1)}), including the 
momentum-conservation constraint, as the proper ``reference'' distribution of 
associated-particle momenta ${{\bf p}_T}_2$, over which dynamical effects are to
be investigated. 
This is admittedly slightly unsatisfactory, since the conditional probability 
distribution is not measured, and depends on two unknown quantities: the total 
number $N$ of final-state particles and the mean square transverse momentum 
$\mean{{\bf p}_T^2}$ (albeit only through their product).\footnote{Furthermore, 
  the averages $\mean{\cdots}$ involve the non-measurable distribution ``in the 
  absence of momentum conservation'' $F({\bf p})$, rather than the physical 
  distribution $f({\bf p})$~\cite{Chajecki:2006hn}.
  This is however but a minor issue, as the difference between $F$ and $f$ is 
  of subleading order in $1/N$, while here we only consider leading-order 
  quantities.} 
Yet they can be estimated, as was done to take into account the conservation of 
total momentum in analyses of anisotropic flow~\cite{Borghini:2000cm,%
  Borghini:2002mv} or in femtoscopy studies~\cite{Chajecki:2006hn}. 
Only at this price can one determine the ``minimally-biased background'' which 
is what the distribution of associated particles would look like in the absence 
of non-trivial correlations. 
For Au-Au collisions at RHIC energies, the corresponding correction might be of 
at most 1 or 2\% when choosing a high-$p_T$ trigger particle; yet this is of 
the same relative magnitude as the effects that are measured~\cite{%
  Adams:2005ph}, so that such a precision is necessary if one wants to 
establish the existence of specific dynamical effects and to quantify their 
importance. 

Let me now comment on the two-particle cumulants in the more general case where 
particles are not emitted isotropically in the plane transverse to the beam. 
More precisely, I shall consider particles emitted with a mean second-harmonic 
transverse anisotropy (elliptic flow) $\bar v_2$ defined by 
$\bar v_2\equiv\mean{p_x^2-p_y^2}/\mean{p_x^2+p_y^2}$.\footnote{This definition 
  yields values of $\bar v_2$ that are typically a factor 2 larger than those 
  obtained with the more conventional definition 
  $v_2\equiv\mean{(p_x^2-p_y^2)/(p_x^2+p_y^2)}$, so that at RHIC energies 
  $\bar v_2\approx 0.1$ in mid-central Au-Au collisions.}
This definition yields at once the identities 
$\mean{p_x^2}=(1+\bar v_2)\mean{p_T^2}/2$ and 
$\mean{p_y^2}=(1-\bar v_2)\mean{p_T^2}/2$, which one can insert in 
equation~(\ref{c2}):
\begin{equation}
\label{c2bis}
\bar f_c({{\bf p}_T}_1, {{\bf p}_T}_2) = -\frac{2}{N\mean{p_T^2}}\left( 
\frac{p_{1,x} p_{2,x}}{1+\bar v_2} + \frac{p_{1,y} p_{2,y}}{1-\bar v_2} \right),
\end{equation}
from which one deduces the conditional probability distribution 
$f({{\bf p}_T}_2 \big| {{\bf p}_T}_1)$ of particles associated to a trigger 
particle with transverse momentum ${{\bf p}_T}_1$. 
One sees that the effect of the constraint from momentum conservation is 
larger in the $y$-direction, i.e.\ perpendicular to the nucleus-nucleus 
impact parameter, than in the $x$-direction. 
That is quite normal, since particles with a transverse momentum along $y$ are 
less numerous than those pointing in the $x$-direction (remember that values of 
$v_2(p_T)=0.16$ were reported for $p_T\geq 2$~GeV$/c$ in minimum-bias Au-Au 
collisions at RHIC~\cite{Adams:2004bi}, meaning that twice more charged hadrons
are emitted in the $x$-direction than perpendicular to it). 
As a consequence, if the trigger particle is chosen with ${{\bf p}_T}_1$ in the 
$y$-direction, only few particles are present to balance its momentum, so that 
those few ones are more strongly correlated to it than if we were considering a 
trigger particle with a momentum in the $x$-direction. 

The dependence of the strength of the two-particle cumulant~(\ref{c2bis}) on 
the azimuths of the two particles means that the correction for the 
momentum-conservation effect has to be performed with some care. 
Let me illustrate that with an example. 
I have already argued above why one should use the conditional probability 
distribution $f({{\bf p}_T}_2 \big| {{\bf p}_T}_1)$ instead of the marginal one 
when investigating the possible structures associated with the presence of a 
(high-$p_T$) trigger particle. 
If only the average correction, corresponding to the isotropic case~(\ref{c2t}),
were used instead of the azimuthally-dependent one~(\ref{c2bis}), this would 
yield an over-correction (resp.\ an under-correction) of 
$f({{\bf p}_T}_2 \big| {{\bf p}_T}_1)$ for trigger particles emitted along the 
nucleus-nucleus impact parameter (resp.\ emitted along the $y$-axis). 
As a result, one would observe a small (using typical RHIC values for $N$, 
$\mean{p_T^2}$ and $\bar v_2$, of order $\lesssim 0.005$) spurious bump (resp.\ 
dip) at 180$^{\rm o}$ away from the trigger in the conditional probability 
distribution for ${{\bf p}_T}_2$.
This azimuthally-dependent spurious structure due to an inaccurate definition 
of the ``minimally-biased background'', which is to be the reference over which 
correlations of dynamical origin can be observed, would prevent any accurate 
determination of these interesting correlations, by mimicking irrelevant 
features. 
Further implications of the azimuthal dependence of the two-particle 
cumulant~(\ref{c2bis}) due to momentum conservation are discussed in 
reference~\cite{Borghini:2007xx}.
\bigskip

In summary, I have recalled that the general purpose of studying correlations 
is to yield evidence of phenomena that go beyond trivial expectations. 
In the specific context of high-energy collisions, correlations between any 
number of final-state particles are induced by the conservation of total 
momentum, which are not of dynamical origin. 
These uninteresting correlations can be computed, thereby allowing one to define
a ``minimally-biased background'', including the effect of total-momentum 
conservation, which is viewed as the reference over which genuine dynamical 
effects might be revealed.


\begin{thebibliography}{99}

\bibitem{Lisa:2005dd}
See e.g.\ M.~A.~Lisa, S.~Pratt, R.~Soltz and U.~Wiedemann,
{\em Femtoscopy in relativistic heavy ion collisions\/},
{\em Ann.\ Rev.\ Nucl.\ Part.\ Sci.\/} {\bf 55} (2005) 357 [nucl-ex/0505014].

\bibitem{Lednicky:1995vk}
R.~Lednick{\'y}, V.~L.~Lyuboshits, B.~Erazmus and D.~Nouais,
{\em How to measure which sort of particles was emitted earlier and which 
  later\/}, 
{\em Phys.\ Lett.\ B\/} {\bf 373} (1996) 30.

\bibitem{Tang:2007kv}
See e.g. A.~Tang, {\em Collective dynamics at RHIC\/}, 
{\em J.\ Phys.\ G: Nucl.\ Part.\ Phys.\/} {\bf 34} (2007) S277 
[nucl-ex/0701041].

\bibitem{Adler:2002tq}
C.~Adler {\it et al.} (STAR Collaboration), 
{\em Disappearance of back-to-back high $p_T$ hadron correlations in central 
  Au+Au collisions at $\sqrt{s_{_{NN}}}=200$~GeV\/},
{\em Phys.\ Rev.\ Lett.\/} {\bf 90} (2003) 082302 [nucl-ex/0210033].

\bibitem{Adams:2005ph}
J.~Adams {\it et al.} (STAR Collaboration),
{\em Distributions of charged hadrons associated with high transverse momentum
  particles in $pp$ and Au+Au collisions at $\sqrt{s_{_{NN}}}=200$~GeV\/},
{\em Phys.\ Rev.\ Lett.\/} {\bf 95} (2005) 152301 [nucl-ex/0501016].

\bibitem{Foster:1973mt}
M.~C.~Foster, D.~Z.~Freedman, S.~Nussinov, J.~Hanlon and R.~S.~Panvini,
{\em Azimuthal correlations of high-energy collision products\/},
{\em Phys.\ Rev.\ D\/} {\bf 6} (1972) 3135.

\bibitem{Pruneau:2006gj}
C.~A.~Pruneau,
{\em Methods for jet studies with three-particle correlations\/},
{\em Phys.\ Rev.\ C\/} {\bf 74} (2006) 064910 [nucl-ex/0608002].

\bibitem{vanKampen}
N.~G.~van Kampen, {\em Stochastic processes in physics and chemistry\/}, 
North-Holland, Amsterdam 1981.

\bibitem{Danielewicz:1987in}
P.~Danielewicz {\it et al.},
{\em Collective motion in nucleus-nucleus collisions at 800 MeV/nucleon\/}, 
{\em Phys.\ Rev.\ C\/} {\bf 38} (1988) 120.

\bibitem{Borghini:2000cm}
N.~Borghini, P.~M.~Dinh and J.~Y.~Ollitrault,
{\em Is the analysis of flow at the CERN Super Proton Synchrotron reliable?\/},
{\em Phys.\ Rev.\ C\/} {\bf 62} (2000) 034902 [nucl-th/0004026].

\bibitem{Borghini:2003ur}
N.~Borghini, {\em Multiparticle correlations from momentum conservation\/},
{\em Eur.\ Phys.\ J.\ C\/} {\bf 30} (2003) 381 [hep-ph/0302139].

\bibitem{Chajecki:2006hn}
Z.~Chaj{\c e}cki and M.~Lisa, 
{\em Global conservation laws and femtoscopy of small systems\/},
{\em Braz. J. Phys.\/} {\bf 37} (2007) 1057 [nucl-th/0612080].

\bibitem{Borghini:2006yk}
N.~Borghini, 
{\em Momentum conservation and correlation analyses in heavy-ion collisions at
  ultrarelativistic energies\/},
{\em Phys.\ Rev.\ C\/} {\bf 75} (2007) 021904 [nucl-th/0612093].

\bibitem{Borghini:2002mv}
N.~Borghini, P.~M.~Dinh, J.~Y.~Ollitrault, A.~M.~Poskanzer and S.~A.~Voloshin,
{\em Effects of momentum conservation on the analysis of anisotropic flow\/},
{\em Phys.\ Rev.\ C\/} {\bf 66} (2002) 014901 [nucl-th/0202013].

\bibitem{Adams:2004bi}
J.~Adams {\it et al.} (STAR Collaboration),
{\em Azimuthal anisotropy in Au+Au collisions at $\sqrt{s_{_{NN}}}=200$~GeV\/},
{\em Phys.\ Rev.\ C\/} {\bf 72} (2005) 014904 [nucl-ex/0409033].

\bibitem{Borghini:2007xx}
N.~Borghini, 
{\em Phase space constraints and statistical jet studies in heavy-ion 
  collisions\/},
arXiv:0710.2588 [nucl-th].

\end{thebibliography}
\end{document}